\documentclass{aa}
\usepackage{graphicx} 
\usepackage{natbib} 
\usepackage{amssymb} 
\newcommand{\citeposs}[1]{\citeauthor{#1}'s \citeyearpar{#1}} 
\newcommand{\mybf}[1]{#1}
\def\aa_{\aap} 
\def\adndt{Atomic Data Nuc.\ Data Tables}
\def\rmp{Rev.\ Mod.\ Phys.}
\begin{document} 
 
\title{The surface carbon and nitrogen abundances in models of
ultra metal-poor stars}

\author{H.~Schlattl\inst{1}, M~Salaris\inst{1}, 
S.~Cassisi\inst{2} \and A.~Weiss\inst{3}} 
 
\institute{Astrophysics Research Institute, Liverpool John Moores 
University, Twelve Quays House, Egerton Wharf, Birkenhead CH41 1LD, 
United Kingdom 
\and 
INAF -- Osservatorio Astronomico Collurania, 
Via Mentore Maggini, 64100 Teramo, 
Italy 
\and 
Max-Planck-Institut f\"ur Astrophysik, 
Karl-Schwarzschild-Str.~1, 85741 Garching, 
Germany} 
 
\offprints{H.~Schlattl,\\ \email{hs@astro.livjm.ac.uk}}

\date{Received; accepted} 
 
\authorrunning{H.~Schlattl et al.} 

\abstract{We investigate whether the observed high number of carbon- and 
nitrogen-enhanced extremely metal-poor stars could 
be explained by peculiar evolutionary properties during the core He flash 
at the tip of the red giant 
branch. For this purpose we compute a series of detailed stellar models
expanding upon our previous work; in particular,
we investigate if during the major He flash the 
penetration of the helium convective zone into the overlying hydrogen-rich 
layers can produce carbon- and nitrogen-rich abundances in  
agreement with current spectroscopic observations. The dependence of 
this phenomenon on selected model input parameters, 
such as initial metallicity and treatment of convection is 
examined in detail. 
} 
\maketitle 
\section{Introduction} 
 
Spectroscopic observations of ultra metal-poor (UMP) stars 
have disclosed that carbon and nitrogen are significantly 
overabundant with respect to Fe, compared to typical halo stars 
which show [C/Fe]$\approx$[N/Fe]$\approx$0.0. 
According to \citet{beers:99} and \citet{rbs:99} 
about 10\% of the stars with [Fe/H]$\leq$$-$2.5 display [C/Fe]$\ge$1.0; this 
fraction rises to about 25\% when [Fe/H]$\leq$$-$3.0. 
Also N appears to be overabundant by the same amount, while 
$\alpha$-elements 
show a ratio with respect to Fe that is typical of halo stars, 
i.e.~[$\alpha$/Fe]$\approx$0.4. 
As discussed, e.g., by \citet{nrb:97}, it is difficult to explain these 
abundance ratios in terms of binary star evolution or ejecta from 
very massive Population III supernovae, which in particular produce few
nitrogen but vast amounts of oxygen \citep{hewoo:02},
so that the origin of the surface abundances of 
UMP stars is still shrouded in mystery. 
 
Concerning the nature of the UMPs, different ideas may be followed, the first
one being that they are true Pop~III stars --- i.e.\ of initially zero
metallicity --- with the observed heavy elements (Fe, etc.)
resulting from pollution of just the envelope by other stars such as
supernovae. 
In two recent papers (\citealt[ Paper~I]{wcss:00}, and 
\citealt[ Paper~II]{scsw:01}) we have thus discussed in detail 
the evolution of low-mass zero-metal stars, including the effect 
of atomic diffusion, mass loss and surface metal pollution. One 
important result  
was that the small entropy barrier between the H- and He-rich 
regions allows the He-flash driven convective zone to penetrate the
overlying H-rich layers at the tip of the red giant branch. 
The consequent inward migration of protons into high-temperature 
regions leads to an H-shell flash, which further increases the 
extension of the central convective region. At the late stages of this 
phase the convective envelope deepens and merges with the inner convective 
zone. The initially metal-free surface is therefore enriched by 
a large amount of matter processed in He- and H-burning reactions. In 
particular, very high C and N abundances follow. 
We denote this process of surface C and N enrichment as HElium Flash 
induced Mixing (HEFM). 

\begin{table*}[t]
\caption{Data for the UMP stars discussed in the text. 
The columns show star name, logarithm of the surface 
gravity, effective temperature, [Fe/H], [C/Fe], [N/Fe], and 
\element[][12]{C}/\element[][13]{C} ratio, respectively. The errors on
$T_\mathrm{eff}$ and log(g) of CS~22892-052 have been estimated by
us.\label{data}}
\begin{minipage}{\textwidth}
\begin{tabular}{llllllc} \hline \hline
star & \multicolumn{1}{c}{log(g)} &
\multicolumn{1}{c}{$T_\mathrm{eff}$} & \multicolumn{1}{c}{[Fe/H]} & \multicolumn{1}{c}{[C/Fe]} & \multicolumn{1}{c}{[N/Fe]}& 
\multicolumn{1}{c}{\element[][12]{C}/\element[][13]{C}}\\ 
\hline 
CS 22892-052\footnote{data from \cite{nrb:97}}&1.5$\pm$0.5&4850$\pm$100&$-$2.97$\pm$0.20&1.10$\pm$0.23&1.0$\pm$0.52&$\gtrsim$10\\ 
CS 22957-027\footnote{C \& N abundances from \cite{nrb:97l},
$\mathrm{[Fe/H]}$, $g$, and $T_\mathrm{eff}$ from  \citet{bmb:98}} &2.25$\pm$1.0&4839$\pm$130&$-$3.43$\pm$0.12&2.20$\pm$0.30&2.0$\pm$0.50&$\approx$10\\ 
CS 22948-027\footnote{data from \cite{ANR:02}}&1.0$\pm$0.3&4600$\pm$100&$-$2.57$\pm$0.23&2.0$\pm$0.18&1.8$\pm$0.24& $\approx$10\\
CS 22949-037\footnote{data from \cite{NRB:01}}&1.7$\pm$0.3&4900$\pm$100& $-$3.79$\pm$0.16&1.05$\pm$0.20&2.7$\pm$0.40&---\\\hline 
\end{tabular} 
\renewcommand{\footnoterule}{\vspace*{-1.5ex}\rule{0pt}{0pt}}
\end{minipage}
\end{table*} 

This scenario --- already suggested by \citet{fii:00} --- could 
potentially explain the anomalous abundance pattern at the surface of 
UMP stars, since via this mechanism high abundances 
of C and N are produced in the post He-flash phases. In Paper~II we
have shown that an initially  
metal-free star of 0.82$\,M_{\odot}$, polluted by 
0.0003$\,M_{\odot}$ of $Z$=0.02 material, undergoes HEFM and reproduces
in its post He-flash phase approximately  
luminosity and surface gravity of two of the best studied UMP 
stars, namely CS~22892-052 and CS~22957-027. 
The stellar mass was chosen such that the age on the RGB of 13.7\,Gyr is 
compatible with the current estimates of the 
age of the universe, and with the age of CS~22892-052 estimated by 
\citet{cpk:99} from nuclear chronology (15.6$\pm$4.6\,Gyr). 
The amount and composition of the polluting material was adjusted such that
the observed surface [Fe/H] abundance ([Fe/H]$\approx$$-$3) was obtained and,
as the result of the HEFM, [C/Fe] and [N/Fe] ratios considerably larger than
zero resulted.  
However, the values of [C/Fe] and [N/Fe] predicted by the 
models are about 2 orders of magnitude higher than the observed ones; 
i.e., we predict [C/Fe]$\approx$[N/Fe]$\approx$4, while observations of these
stars yield [C/Fe]$\approx$[N/Fe]$\approx$1--2 (see Table~\ref{data}). We
note, however, that higher N overabundances are observed, e.g., 
in CS~22949-037 \citep{NRB:01}. 

An alternative explanation for the C and N enhancement in UMPs \mybf{is
provided by HEFM} during the thermally pulsating asymptotic giant-branch
phase (TP-AGB) which occurs in metal-free stars with
$M$$\ga$1$\,M_\odot$~\citep{CCT,fii:00}. However, age, surface gravity,
and effective temperature of these AGB stars are more difficult to reconcile
with those of the observed objects, which are therefore unlikely to be TP-AGB
stars themselves. The idea in this case is that the observed
low-mass UMP objects form binary systems with more massive AGB stars, which have
transferred part of their C- and N-enriched 
envelope to their companion~\citep{fii:00}. 
A potential problem of this
hypothesis is that HEFM during the TP-AGB of intermediate-mass stars may 
enhance the surface oxygen content similar to carbon. \mybf{This feature
can be found in models of \citet{SLL:02}, which however do not agree
in this respect with
computations of \citet{CDLS:01}}.
An oxygen enhancement of [O/Fe]$\approx$1--2 might
not necessarily be in
contradiction to observations, as only very weak upper limits on
the [O/Fe] ratio exist \citep{smcp:96,nrb:97}. 
\mybf{But} the main problem of this scenario is that not all UMPs are in
binary systems, and the orbital periods of those who are, are
inconsistent with an AGB mass transfer paradigm\ \citep{PS:02}. Moreover,
the origin of Fe in these stars is as uncertain as in our favoured
scenario. It is presently not clear whether intermediate-mass stars
with [Fe/H]$\approx$$-$3 undergo HEFM during the TP-AGB or rather evolve
like ordinary Pop~II objects. In the latter case the surface C, N, O, and Fe
abundances of UMP stars would then be the result of mass transfer from their
companion and the accretion of further material, e.g., a nearby
supernova. 

In summary, HEFM during the major He flash is presently the most promising
astrophysical explanation for the high [C,N/Fe] ratios observed in UMP
stars despite their failure of reproducing the absolute value of 
the C and N enrichment.
Therefore we are studying to what extend HEFM occurs also in 
low-mass stars with initial $Z$$>$0, and whether those objects are
showing
the observed [C,N/Fe] ratios of CS~22892-052 and CS~22957-027. 
This extension of our model assumption is warranted by the results about star
formation in primordial environments, which agree on the fact that the
formation of (very) massive objects is strongly favoured in the absence of
metals \citep{abn:02,bcl:02}. If this indeed is the case, the UMPs would
constitute the extreme end of Pop~II, and had formed out of material already
bearing the signature of individual supernova events
\citep{shitsu:98}.

\begin{figure*} 
\sidecaption
\includegraphics[width=12cm]{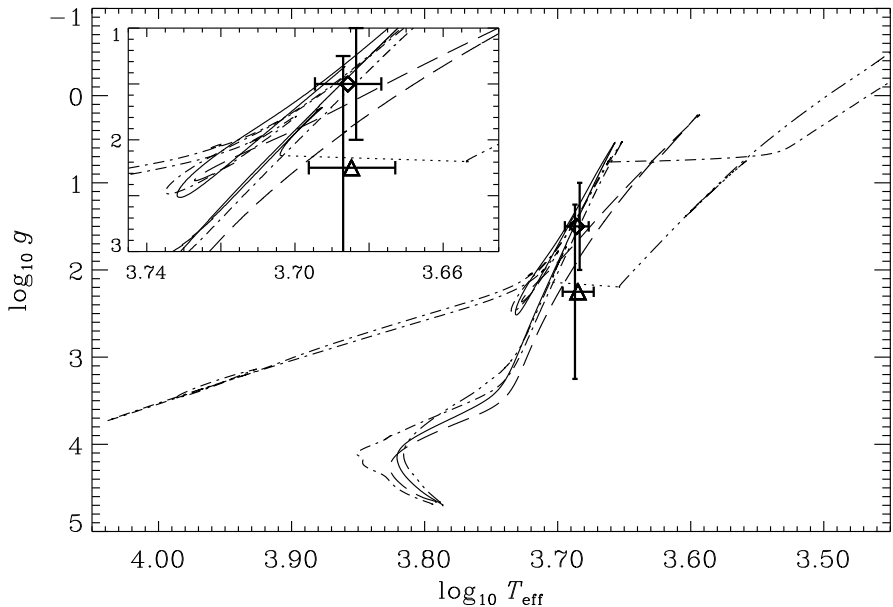}
\caption{\ Evolution in the gravity-effective temperature 
diagram of models with different initial metallicities: 
a $M$=0.82$\,M_{\odot}$-models with  
initial $Z$=0 and surface metal pollution (dash-dot-dot-dotted line), a 
0.78$\,M_{\odot}$-model with $Z$=$2.0\times10^{-5}$ (`best  
model', solid line),  
and a 0.82$\,M_{\odot}$-model with $Z$=$6.0\times10^{-4}$, where  
an initial \mbox{C-,} N-enhanced metal-mixture has been assumed 
(dashed line). In addition, a model with $Z$=$2.0\times10^{-5}$ and
additional overshooting ($F=0.03$) undergoing a HEFM during the
TP-AGB phase is shown (dash-dotted line). Open symbols show the
position of the two UMP stars CS 22892-052 and CS 22957-027
(Table~\ref{data}). The vertical error bars 
are slightly offset in order to not overlap.\label{hr1}}
\end{figure*} 

The final alternative for the origin of the carbon-rich UMPs is
that they were already born with the abundances they now show. This scenario
would receive support if single dwarfs of similar composition should
definitely be identified. While
one shifts the problem of identifying the source of the peculiar abundances to
other stars, it is worthwhile, nevertheless, to investigate the evolution of
such C-rich stars. For example, they are expected to reach standard RGB-tip
luminosities and low gravities, which could be compared with those of observed
objects. Stars of extremely low metallicity ($Z=0$ in particular), have a
definitely shorter RGB, in contrast.

In addition to the three plausible models for the origin of the C-rich UMPs,
we also investigate in this paper a physical problem associated with the
HEFM. 
Due to the large uncertainties still affecting the treatment of convection
in stellar models, we will assess the sensitivity of the HEFM process to
different assumptions about the extension and mixing efficiency of
convective regions in the 
stellar interiors.

The paper is structured as follows. Section~2 summarizes briefly the 
model computational techniques, input physics, and evolution of 
low-mass stars undergoing HEFM. 
Section~3 discusses the occurrence of HEFM and resulting surface 
abundances for different initial metallicities and various
parameterizations of the stellar convective regions;
Sect.~4 summarizes the results.

\section[]{Modelling and evolution of stars undergoing HEFM during the
He flash} 
 
All calculations (as the ones in Paper~II) have been performed 
using the Garching stellar evolution code (\citealt{wsch:2000}). 
The numerical details are presented in Paper~II and will not be 
repeated here. The important feature of our code is the ability to
follow in detail the evolution through the He 
flash considering 
simultaneously mixing and burning processes. Mixing 
is described by means of a time-dependent algorithm 
that treats convection as a fast diffusive process, where the 
diffusion constant is proportional to the convective velocity obtained 
from a convection theory. We have discussed in Paper~II that 
the convective velocities predicted by both mixing-length and 
\citeposs{cm:91} theory lead to very similar results. 
In general, arbitrary changes by up to 2 orders of magnitude in the 
mixing speed do not affect 
appreciably the HEFM process. 

Atomic diffusion is treated according to \citet{tbl:94}, 
opacities are from \citet{ir:96} and \citet{af:94}, neutrino 
energy losses from \citet{mki:85}, and the equation of state (EOS) 
is a Saha EOS plus a 
simplified description of the degenerate electron gas in the core 
regions following \citet{kw:90}. 
Our reaction network follows the abundances of H, \element[][3]{He}, 
\element[][4]{He}, \element[][12]{C}, \element[][13]{C}, 
\element[][14]{N}, \element[][15]{N}, \element[][16]{O}, 
\element[][17]{O}, without assuming a priori equilibrium compositions 
for any of these chemical elements. The reaction rates for the
$pp$- and CNO-burning have been taken from \cite{Adel:98}, while the
$\element[][12]C(\alpha,\gamma)\element[][16]O$-rate agrees with
\citeposs{CF:85} value. The latter one is, in the relevant temperature
range ($0.1 < T/(10^9\,\mathrm{K}) < 0.2$),  about a
factor of 2 higher than the rates of \citet{CF:88} and about 30--50\%
higher than the most recent value of \citet{KFJ:02}.

Figure~\ref{hr1}, as a reference, displays the evolutionary track of 
the model with 0.82$\,M_{\odot}$, $Z$=0, and surface chemical 
pollution mentioned in the introduction (similar to model B1 of Paper
II), which experiences the HEFM.

The 0.0003$\,M_{\odot}$ of $Z$=0.02 polluting material are instantaneously 
accreted before the star reaches the zero-age main sequence; this 
approach 
maximizes the effect of accretion and diffusion effects. After this 
episode the star evolves along the main sequence producing energy 
through the $pp$-chain. Diffusion is never able to bring the surface 
metals deep enough to contribute to the burning.  
 
At the sub-giant branch 
the core increases its temperature and density, and gradually 3-$\alpha$ 
reactions are setting in. In stars more massive than $\approx0.9\,M_\odot$ the 
central H content is not fully exhausted when sufficient C is produced 
($X(\element[][12]C) \approx 10^{-10}$) 
to gain considerable nuclear energy from the CNO-cycle. As a consequence a 
thermal runaway through the  CNO-cycle ensues, causing  
a characteristic blue loop in the H-R diagram. The consequent 
expansion 
of the inner regions reduces the efficiency of the  3-$\alpha$ process 
and CNO-cycle, terminating the runaway which lasts only about $10^7$ 
years. In less massive models, as the ones 
considered in this work, the central H is already 
exhausted before sufficient C can be produced, and thus no blue-loop 
occurs on the sub-giant branch. 
 
Regardless of the occurrence of this CNO flash, all low-mass 
models settle 
on the RGB and evolve towards higher luminosities. While on the
sub-giant branch their main energy source 
remains the $pp$-chain, the contribution of the CNO-cycle is
increasing gradually with luminosity as more carbon is created at
the inner tail of the H-burning shell by 3-$\alpha$ reactions. In a
0.8$\,M_\odot$ star about 50\% of the H-burning energy 
star is produced by the CNO-cycle at the tip of the RGB. 

\begin{figure}
\resizebox{\hsize}{!}{\includegraphics{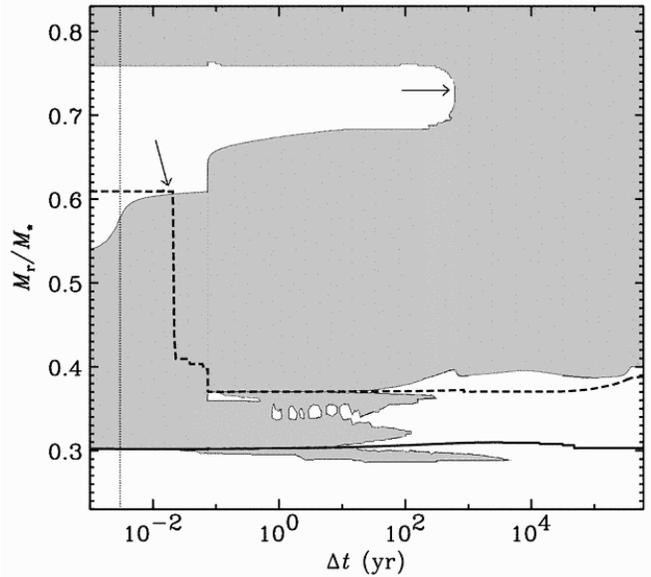}}
\caption{Evolution of the different convective zones --- indicated 
by shaded areas --- from the He-flash 
ignition onwards in the $Z$=0 model of Fig.~\ref{hr1} with surface 
pollution. The dashed and solid lines represent 
the mass shells (in units of fraction of total mass) 
of maximum energy release by 
H and He burning, respectively. The two arrows mark the penetration of
the He-flash driven convective zone into H-rich layers and the
merger of the convective envelope with the C- and N-rich
inner convective region, respectively. The moment of maximum
He-fusion energy release is indicated by the vertical dotted line.
The zero point in time has been chosen to be 1~day before this
moment. The short-living small radiative layers between the shells 
are only marginally stable against convection (``semi-convection'').
\label{conv1}} 
\end{figure}

At this stage the mass difference between 
the location of the maximum energy generation in the core and the 
He-core boundary is $\approx$0.29$\,M_{\odot}$. 
Soon after the onset of the He flash a 
convective shell develops, which reaches the H-rich matter in the 
envelope about 1 month after the flash (left arrow in Fig.~\ref{conv1}); 
as soon as H is carried into the interior hotter convective region, it 
starts to burn at a very high rate. Due to this extra energy input 
the upper boundary of the 
convective shell immediately moves closer to the surface 
(thus ingesting even more protons), while the He-burning rate is 
significantly reduced. The single inner convective region 
splits into two zones, one for the He- and one for the H-burning. 
The rapidly weakening He-burning shell can no longer support the 
underlying layers against contraction, and the released gravothermal
energy forms a further convective region which disappears again after a few 
1,000 years\footnote{This convective
region appeared also in the calculations presented in Paper~II, 
but was not shown in the corresponding figure due do
insufficient plot data.}. 

About 500\,yr after the H ingestion the 
convective envelope deepens and merges with the convective region 
above the H-burning region (horizontal arrow in Fig.~\ref{conv1}). A huge amount of matter processed through 
H and He burning is brought to the surface enriched 
in He, C (produced during the He burning) and N (produced by the 
CNO cycle at the expenses of C). 
The resulting surface [C/Fe] and [N/Fe] ratios are, respectively, 4.1
and 4.3, while the surface oxygen abundance is practically unaltered
with respect to the initial value (Table~\ref{summ}).
 
As a result of this dredge up of heavy elements 
(which happens on timescales of a few weeks) the envelope
opacity is increased significantly; this produces an abrupt discontinuity in the 
effective temperature visible in Fig.~\ref{hr1}. 
At this stage, with a practically ceased He burning, the star behaves 
as a newborn RGB star, climbing its own RGB. The H shell produces the 
energy needed to support the star. Another He flash ensues at the end 
of this second RGB phase, this time with a smaller and less degenerate 
core. The flash is weaker and no HEFM happens. 
The subsequent evolution does not show any peculiar feature. 

\begin{figure} 
\resizebox{\hsize}{!}{\includegraphics{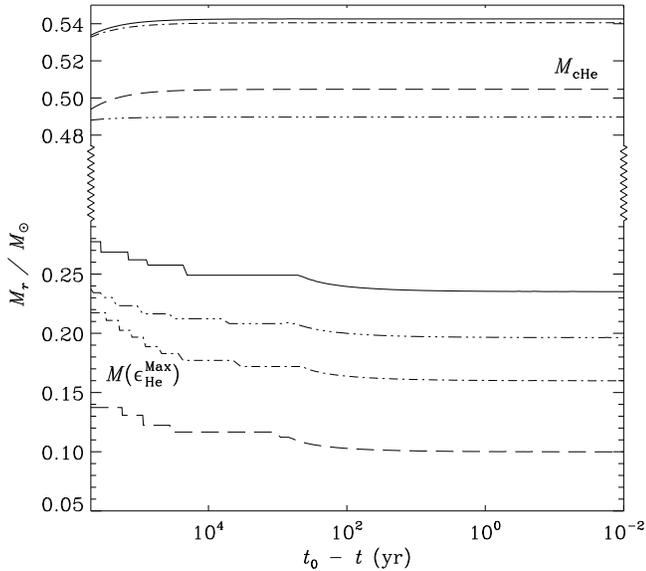}}
\caption{Mass location of the He-core upper boundary 
and energy generation maximum as a function of the time 
before the flash ($t_0$ indicates the onset of the flash), 
for the same models as in Fig.~\ref{hr1}, indicated by the same
line styles.\label{core}}
\end{figure} 
 
As a general rule, all the changes in the 
physical inputs and/or initial conditions that can contribute to an 
increase of the electron degeneracy and thus cause the location of the 
He-flash ignition to move closer to the border of the He core favour 
the occurrence of HEFM. Changes in the relative location 
as small as $\approx$0.02$\,M_{\odot}$ can make a difference between 
models undergoing HEFM and models which do not experience it. 
In Paper~II we have shown how 
an increase of the initial He abundance, or an increase of the initial 
mass, or the inclusion of heavy element pollution at the stellar 
surface are disfavouring the onset of HEFM. On the other hand, the 
inclusion of element diffusion favours this process, while the 
inclusion of mass loss from the stellar surface does not affect at all 
the 
onset of the HEFM. 
 
\section[]{The influence of initial metallicity and convection} 
 
In Paper II we have investigated the dependence of the HEFM process on 
several input-physics parameters and various 
assumptions adopted in computing stellar models; however,  
we have considered only stars with initial zero metallicity,
plus some eventual metal pollution at the surface.
It is worthwhile to investigate the dependence of the HEFM 
on the value of the initial stellar metallicity, given the unknown composition
of these stars at the time of formation, as outlined in the introduction.

Because of the uncertainty in the treatment of convection  
in stellar structures, it is also important to verify how much the
HEFM process is sensitive to 
different assumptions about the extension and mixing efficiency of
convective regions in the 
stellar interior. In particular, we considered various amounts of
overshooting from the canonical formal convective boundaries fixed by the
Schwarzschild criterion.

In this section we will discuss these effects 
in theoretical models representative of 
CS~22892-052 and CS~22957-027. 
 
\subsection[]{Models with $Z$$>$0} 
 
Models with initial $Y$=0.23 and increasing $Z$ have been computed 
in order to derive the value of the maximum initial metallicity which 
produces HEFM in a star with 0.82$\,M_{\odot}$. The heavy element 
mixture has been considered to be $\alpha$-enhanced, with 
[$\alpha$/Fe]=0.4; carbon and nitrogen were scaled solar.
We obtain HEFM only for $Z<10^{-7}$, corresponding to [Fe/H]$<$$-$5.6. 
Since all observed [Fe/H]-abundances in UMPs so far are higher than $-4$, this
already excludes the possibility that the C/N-anomalies were produced in stars
with initial Fe-abundances as observed today.
Due to 
the mixing of H-rich material into the 
H-depleted interior during the HEFM,  
the surface H abundance drops by about a factor of two, 
and the upper limit for the surface Fe abundance after this phase is 
[Fe/H]$<$$-$5.3, about 
2 orders of magnitude smaller than the observed metallicities of 
CS~22892-052 and CS~22957-027. 
Whenever HEFM happens, the [C/Fe] and [N/Fe] ratios in the post 
He-flash phase are always of the order of 4, still about 2 orders of 
magnitude higher than the observed values. 

This upper value for the metallicity is slightly increased if we take 
into account the revised plasma-neutrino rates 
by \citet{hrw:94}. In this case the limiting metallicity is $Z=10^{-6}$, 
corresponding to [Fe/H]=$-$4.3 after the HEFM, i.e., 1\,dex higher than 
in case of the old neutrino losses. The reason is that the  
degeneracy of the core increases and the He flash starts more 
off-centre.  
We also wish to notice here that, in spite of their different input 
physics 
and cruder treatment of the coupling between convection and burning, 
\citet{fii:00} found a similar upper limit for the metallicity 
of a 0.80$\,M_{\odot}$ star undergoing HEFM, 
and similar values of the surface [C,N/Fe] ratios.
 
We have also tested the effect of implementing a new EOS by 
\citet{i:2002}, which covers the entire stellar structure in all 
relevant evolutionary phases, and closely reproduces the OPAL EOS 
\citep{rsi:96} in the common validity range \citep[see, e.g.,][]{s:2002}. 
The inclusion of this new EOS in the models 
does not modify the value of the largest critical heavy elements 
abundance for which the HEFM occurs. However, with this equation of
state the stellar ages are 
slightly reduced (by about 0.5\,Gyr) with respect to our reference EOS, 
and therefore 
one has to choose a slightly lower mass to reproduce the observations, 
which might favour the occurrence of HEFM. 

Therefore, we have computed a, what we call, `best model' with an 
initial [Fe/H] equal to the observed one, and a mass  
which yields an age at the RGB tip 
well matching the age estimated for CS~22892-052 and CS~22957-027; we 
considered 
$M$=0.78$\,M_{\odot}$, $Z$=$2.0\times10^{-5}$ (which corresponds, including the $\alpha$ 
enhancement by 0.4\,dex, to [Fe/H]=$-$3.3), neutrino emission following
\citet{hrw:94}, \citeposs{i:2002}  
EOS, plus atomic diffusion of H, He, and metals (which further 
reduces the ages by about 0.5\,Gyr). The corresponding evolutionary track is 
shown in Fig.~\ref{hr1}, too (solid line). In spite of the reduced
stellar mass  the effect of the initial 
metallicity dominates, and \emph{no} HEFM occurs. The location of the maximum 
energy generation at the flash is 0.30$\,M_{\odot}$ away from 
the boundary of the He core (solid line in Fig.~\ref{core}), a
distance about 0.01$\,M_{\odot}$ larger than in the cases when HEFM
occurs. We computed also the evolution of stars of 0.75$\,M_{\odot}$ and 
0.82$\,M_{\odot}$ with the same initial composition as the
0.78$\,M_{\odot}$ star 
to span the age range determined for CS~22892-052 
from nuclear chronology, not obtaining HEFM in any case. 
 
\subsection[]{Models including overshooting}\label{over}
 
Models including overshooting have been computed in order to check its 
influence on both the onset of HEFM and the amount of C and N dredged up 
to the surface. Since overshooting increases the 
extension of convective regions, the occurrence of HEFM is in 
principle favoured. 
Our overshooting description follows \citet{bhf:98} and 
it is modelled as an exponential diffusive process. The diffusion 
constant of the overshoot region decays exponentially 
outside the Schwarzschild convective boundary, starting from the value 
assumed by the convective diffusion coefficient at the convective 
boundary; 
the decay length is $F$ times the pressure scale height at the convective 
boundary. 
Hydrodynamical simulations of shallow 
convective envelopes in A stars by \citet{fls:96} predict 
$F$=0.25$\pm$0.05, while for the 
overshooting from the convective envelopes of AGB stars \citet{hbse:97} took
$F=0.02$.

We have computed a series of models
(adopting the same input physics as in Paper II)
with different values for $F$, 
considering 
$M$=0.82$\,M_{\odot}$, $Y$=0.23 and $Z$=$2.0\times10^{-5}$ which corresponds to 
[Fe/H]=$-$3.3, approximately the [Fe/H] 
value determined for the two UMP stars under scrutiny.
In these models the simpler EOS (Saha-EOS completed with degenerate
electron gas in the deep interior) and the neutrino
losses of \cite{mki:85} have been used.
 
We have first considered overshooting from 
all the convective boundaries, starting \mybf{from 
the main-sequence} phase. As a general result, we found that 
the inclusion of overshooting during the main sequence increases the 
lifetime of the small convective core which develops at the beginning 
of the main sequence, when the \element[][3]{He} abundance gradually 
reaches nuclear   
equilibrium in the centre; this, in turn, 
causes an increase of the main-sequence lifetime and a decreased 
electron degeneracy in the He core along the RGB, which disfavours the 
occurrence of HEFM. Moreover, the increased lifetime would 
demand a larger stellar mass in order to obtain not too 
high stellar ages, and that would further reduce the probability for HEFM.  
 
Models with $M$=0.82$\,M_{\odot}$ and F$=$0.03 do not show HEFM, and are 
continuing their evolution through the horizontal branch to the
TP-AGB (dash-dotted line in
Fig.~\ref{hr1}). During this phase the
inter-shell convective zone is able  
to penetrate the H-rich envelope, similar to the HEFM
process~\citep{CCT,fii:00}. The surface C and N abundances
rise again by about 4\,dex, while O is increased by about
3\,dex. Hence, an O enrichment about one order of magnitude smaller
than for C is expected in these stars, which implies that
[O/Fe] should be at most~1 for the stars in Table~\ref{data}.
Since no oxygen abundance could be determined in
these stars, but only an upper limit of [O/Fe]$\la$0.6 \citep[][for
CS~22982-052]{smcp:96}, this scenario could provide a
possible explanation for C-rich UMPs, too.
However, surface gravity and temperature of these AGB stars are
too low to be in 
agreement with the stars under scrutiny (Fig.~\ref{hr1}).

Higher values of $F$ do not lead to HEFM. In fact, for $F=0.1$ there is no 
He flash, but quiescent He-burning ignition and of course no HEFM. 
 
In case overshooting is included only from the RGB phase onwards, 
the stellar lifetime is not 
affected, and one needs at least $F=0.17$ to get HEFM. This value is about
30\% smaller as the one obtained by \citet{fls:96} for A-star
envelopes, but appears to be quite high compared to that needed in AGB
stars
\citep{hbse:97}; moreover, switching on 
overshooting only during the RGB phase appears to be entirely ad hoc; 
in addition, the 
resulting [C/Fe] and [N/Fe] ratios after the HEFM are again of the 
order of 4, i.e., too high with respect to the observations. 

\subsection[]{Models with reduced mixing efficiency}\label{mixeff}

{From} the computations performed until now, it appears
that one of the main problem to the HEFM scenario is the too high C
and N enrichment of the surface. 
One solution to obtain lower surface [C,N/Fe] 
ratios would be to have a smaller overlap between the convective region 
developing at the flash, and the overlying H-rich region. 
Smaller overlap implies less protons and therefore less processed C and 
N.  This smaller overlap could be achieved by strongly reducing  
the diffusion coefficient associated to the convective mixing. 
Using standard mixing-length velocities, the crossing time of the 
convective shell is of the order of a few hours. 
A reduction of the ``convective'' diffusion coefficient
is performed by simply reducing the multiplicative factor that links 
the coefficient itself to the convective velocity. 
 
The upper boundary of the He-flash driven 
convective shell is at first order dictated by the energy of the flash 
and the location of the ignition point. However, after some protons 
are engulfed the increased energy production (due to the additional
H burning) pushes the upper boundary 
further up, thus increasing the number of proton ingested. 
If the mixing efficiency is reduced, protons are mixed less deep and 
less energy is produced, with a smaller increase of the convective 
shell extension. This would cause a smaller final number of protons 
ingested and a lower N production. 
 
\begin{figure}[t] 
\resizebox{\hsize}{!}{\includegraphics{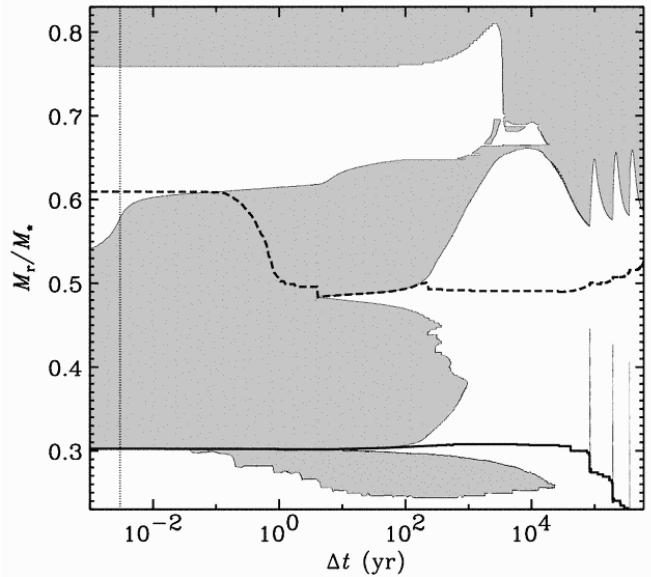}}
\caption{Same model as in Fig.~\ref{hr1} but reducing the convective
mixing efficiency by  
a factor $2\times10^4$.\label{conv2}}
\end{figure} 
 
In order to explore 
this scenario in more detail, we recomputed the
polluted $Z$=0 model of Sect.~2, which 
experiences HEFM, decreasing the ``convective'' diffusion coefficient 
by a factor of $2\times10^4$. The extension of the convective regions are 
displayed in Fig.~\ref{conv2}. 
The upper boundary of the H-flash driven convective shell extends less
into H-rich layers, and the 
location of the H-burning region is less deep; 
the thickness of the H convective shell before 
the dredge up is also thinner, implying a reduction of the 
amount of C dredged up with respect to the case with standard convective 
mixing. The final surface abundance ratios [C/Fe] and [N/Fe] are in this 
case 2.3 and 3.2 (Table~\ref{summ}), respectively, closer to the observed
values.  

The surface 
\element[][12]{C}/\element[][13]{C} ratio is also affected, 
being now increased to 
5.6 compared to 4.5 which we obtain by using 
the standard mixing efficiency. This increased value is in slightly 
better agreement with the observed values of about 10. 
In spite of the much reduced mixing efficiency the dredge up 
still happens fast, on timescales of the order of $10^3$ years. 

\section[]{Discussion} 
 
The results shown in the previous section highlight the fact 
that it is very difficult to reproduce the surface abundances of 
CS~22892-052 and CS~22957-027 from single-star evolution, at least in the
case the observed surface abundances pattern is not  
primordial. To investigate this alternative,
we have computed additionally the evolution of an 0.82$\,M_{\odot}$ star, 
with the same physics of our `best model' and $Y$=0.23, but with
$Z$=$6\times10^{-4}$ and [C/Fe]=[N/Fe]=2 (implying [Fe/H]=$-$3.3),
in agreement with the observed abundances. The
evolutionary track is  
shown in Fig.~\ref{hr1}, while the He-core boundary before and at
He ignition is plotted in Fig.~\ref{core}. As expected, this model does not 
experience HEFM (the He flash starts very deep in the core), so that 
the surface [C/Fe] and [N/Fe] ratios are not modified with respect to 
the initial values. Note that the effective temperature along the
RGB is higher than for the case with HEFM, because of the much lower
C-abundance. The location of the track fits reasonably well 
the observed gravities and effective temperatures, and thus, this
model would be able to reproduce the observed properties of CS~22957-027.
However, as discussed in Sect.~1, if this were a realistic model for the
observed object, the source for the high carbon and nitrogen abundances still
remains unknown, as it is not clear how to pollute 
the interstellar medium with matter showing these anomalous abundance
ratios.

\begin{table}[t]
\caption{Surface abundances of initial metal-free 0.82$\,M_\odot$ stars
polluted by $3\times10^{-4}$$\,M_\odot$ of alpha-enhanced
([$\alpha$/Fe]=0.4]) material with $Z=0.02$ at the ZAMS. All models
contain updated neutrino-losses~\citep{hrw:94} and \citeposs{i:2002}
EOS. The evolution of the convective
regions after the flash are shown in the figures denoted in the first
column. $f_D$ is the factor by which the standard diffusion efficiency
has been multiplied in each model.\label{summ}}
\begin{tabular}{cccccccc}\hline\hline
Fig. & $f_D$ & $[\frac{\element[][]{C}}{\element[][]{Fe}}]$ & 
$[\frac{\element[][]{N}}{\element[][]{Fe}}]$ & 
$[\frac{\element[][]{O}}{\element[][]{Fe}}]$ & 
$[\frac{\element[][]{Fe}}{\element[][]{H}}]$ &
$\frac{\element[][12]{C}}{\element[][13]{C}}$ &
$\frac{\element[][13]{C}}{\element[][14]{N}}$ \\ \hline 
\ref{conv1} & 1 &  4.1 & 4.3 & 0.7 & -3.4 & 4.5 & 0.36 \\
\ref{conv2} & 5$\times$$10^{-5}$ & 2.3 & 3.2 & 0.4 & -3.4 & 5.6 & 0.07 \\ \hline
\end{tabular}
\end{table} 

Thus, assuming that the observed surface composition is not primordial, 
the HEFM is presently the only way to get a surface C and N 
enhancement without invoking some kind of mass transfer from binary 
companions. 
{From} our tests in Sect.~2 there appear to be two main problems with 
the HEFM scenario:
 
Firstly, too much of C and N is transported to 
the surface after the HEFM, and secondly, no HEFM appears to occur for
an initial metallicity that can match the observed [Fe/H] of the stars under
scrutiny, unless surface heavy element pollution has been efficient,
as discussed in \mbox{Paper II}.

Once again, we want to point out that, in spite of their different input 
physics 
and cruder treatment of the coupling between convection and burning, 
\citet{fii:00} found similar upper limits for the metallicity 
of stars undergoing HEFM, and similar values of the [C,N/Fe] ratio for 
stellar mass and metallicity values similar to our `best model'. 
 
Concerning the appearance of HEFM in stars with [Fe/H] in agreement
with observations, one possible solution would be, as discussed in
Sect.\ref{mixeff}, a completely ad hoc choice of overshooting efficiency. 
Another possibility is related to the H depletion during the dredge up 
of the HEFM products. During this phase the 
convective envelope reaches deeper regions involved in the HEFM, which 
are devoid of H. According to our models the surface [Fe/H] increases by 
$\approx$0.3\,dex because of the dilution of H. If the drop in
hydrogen abundance is about five
times higher, a surface [Fe/H] of about $-$3.3 could be obtained
with an initial [Fe/H]=$-$4.3 corresponding to the upper limit of
getting HEFM. 

In order to achieve a larger H depletion, 
a smaller envelope thickness (in mass) is needed, i.e., a smaller 
H reservoir. We estimate that right before the flash ignition 
an envelope of only about 0.03$\,M_{\odot}$ is needed in order 
to reproduce the observed post-HEFM [Fe/H] surface 
values with our 'best model'; with this envelope mass the star would
still be able to ignite He close to its RGB location (see, 
e.g., \citealt{cc:93}), i.e., the effective temperature and gravity
after the HEFM would still be in accordance with CS~22892-052 and
CS~22957-027. Taking into account the age of the star of about
14\,Gyr, which fixes its initial mass to be about 0.8$\,M_\odot$,
it would imply 
that about 0.27$\,M_{\odot}$ are lost during the evolution, a value 
which is not much higher  
than the average mass loss experienced by globular cluster stars 
(about 0.20$\,M_{\odot}$). The lack of metals 
in the outer layers of UMPs, however, should cause a strong decrease of
mass-loss efficiency in comparison with more metal-rich stars, if
radiative driven winds are the main source of mass loss.
Nevertheless, due to our poor 
knowledge about mass-loss processes in RGB stars, this possibility 
cannot be completely ruled out. 

\acknowledgement{
We warmly thank A.~Irwin for allowing us to use his new equation of 
state as well as for 
help in managing his routines. We also thank him for 
discussions and suggestions about the use of his EOS in the evolutionary 
code. We are grateful to T.~Beers for his continuing interest and support of
this work. One of us (S.C.) acknowledges the financial support by MURST/Cofin2000 
under the project  "Stellar Observables of Cosmological Relevance". 
H.S.\ has been supported by a Marie Curie
Fellowship of the European Community programme ``Human Potential''
under contract number HPMF-CT-2000-00951. 
A.W.\ thanks the Institute for Nuclear Theory
at the University of Washington for its hospitality and the Department of
Energy for partial support during the completion of this work.}

% 
%\bibliographystyle{aa} 
%\bibliography{mpoor} 
\newcommand{\singlet}[1]{#1}

\end{document}